\documentclass[twocolumn,preprintnumbers,amsmath,amssymb,floatfix]{revtex4}
\usepackage{graphicx}
\usepackage{dcolumn}
\usepackage{bm}

\begin{document}
\title{Self-organized pattern formation and noise-induced control from particle computation} 
\author{Thimo Rohlf}
\author{Stefan Bornholdt}
\affiliation{IZBI (Interdisciplinary Center for Bioinformatics), University of Leipzig, 
Kreuzstra\ss e 7b, D-04102 Leipzig, Germany}
\date{\today}

\begin{abstract}
We propose a new non-equilibrium model for spatial pattern formation on the basis of 
local information transfer. Unlike standard models of pattern formation it is not based 
on the Turing instability. Information is transmitted through the system via particle-like 
excitations whose collective dynamics result in pattern formation and control. Here, 
a simple problem of domain formation is addressed by this model in an implementation 
as stochastic cellular automata. One observes stable pattern formation, even in the 
presence of noise and cell flow. Noise stabilizes the system through the production 
of quasi-particles that control the position of the domain boundary. Self-organized 
boundaries become sharp for large system with fluctuations vanishing with a power 
of the system size. Pattern proportions are scale-independent with system size. 
Pattern formation is stable over large parameter ranges with two phase transitions 
occurring at vanishing noise and increased cell flow.  
\end{abstract}
\maketitle

An astonishing property of the development of multicellular organisms is its extreme 
error tolerance and robustness against perturbations \cite{Pearson}. 
A key mechanism that coordinates structure formation during development is the 
self-organization of spatiotemporal patterns of gene activity \cite{Gilbert97}. 
While detailed dynamical models of the involved gene regulation processes are not 
within close reach (mostly due to lack of genomic details), phenomenological models 
of developmental processes have been studied for quite a while. One standard model 
for pattern formation in development is diffusion-driven pattern formation exploiting the 
Turing instability \cite{Turing52}. This principle has been applied to modeling biological 
organisms \cite{GiererMeinhardt72,MeinhardtGierer2000} and is able to account for a 
number of observed features of developmental processes, e.g., in the fresh water polyp 
{\em Hydra} \cite{Meinhardt93}. 

However, diffusion based models, dating from the pre-genomic era, are by definition limited.
When applying this type of model to an organism as well known as {\em Hydra}, for example, 
several unresolved problems persist. E.g., morphogen molecules postulated by the model 
still have not been identified. Further, parameter fine-tuning is needed, 
including a non-trivial hierarchy (separation by orders of magnitude) of diffusion 
constants \cite{LanderNie2002}. Most importantly, several experiments point at specific 
developmental features that are not easily captured by these models as, e.g., 
the reorganisation of the body pattern from a fully random cell assembly \cite{TechnauHolstein} 
and the extreme sharpness and vertical regulation of expression boundaries 
\cite{Bosch}, including scale-invariant proportions. While a chemical 
gradient along the body axis could in principle provide position information, 
the precision of ``gradient readout'' is low at typical (i.e.\ low) morphogen concentrations 
\cite{GurdonBourillot2001}. This indicates that the information processing in the gene 
regulation machinery during development is not accurately captured in the diffusion-based 
picture. Indeed, recent experiments show that a large number of regulatory 
genes are involved in cell differentiation and pattern formation \cite{BoschDevelGenes}. 
It may therefore be time to think about alternatives to diffusion-driven pattern formation, 
and in particular explore the possibilities of information-flow-driven pattern formation.  
The aim of this paper is to contribute to this discussion. 

Let us first recapitulate a simple developmental problem, then define a simple stochastic 
cellular automata model that solves this pattern formation task. It is first demonstrated that 
this system performs {\em de novo} pattern formation, 
independent of initial conditions. Then stability of pattern formation is studied in the 
presence of noise and cell flow. 

Position-dependent gene activation is a frequently observed mechanism in animal development.
One example is the fresh water 
polyp \emph{Hydra}, which has three distinct body regions - a head with mouth and 
tentacles, a body column and a foot region. The positions of these regions are accurately 
regulated along the body axis. In addition, new cells continuously move from the central 
body region along the body axis towards the top and bottom, and differentiate into the respective
cell types according to their position on the head-foot axis. This cell flow requires considerable 
robustness of the regulatory processes. The two remarkable features of this regulation that we 
focus on are the scale-independent position regulation and the "reboot"-like \emph{de novo} pattern 
formation from random initial conditions. 

 Let us consider the simplified problem of regulating one domain, say the foot region 
versus the rest of the body. We consider this as a one dimensional problem
as suggested by the well-defined head-foot-axis in \emph{Hydra}. 
Fig.\ 1 formulates the target pattern of this problem. 
\begin{figure}[tbp]
\let\picnaturalsize=N
\def\picsize{85mm}
\def\picfilename{scalnew.eps}
\ifx\nopictures Y\else{\ifx\epsfloaded Y\else\input epsf \fi
\let\epsfloaded=Y
\centerline{\ifx\picnaturalsize N\epsfxsize \picsize\fi
\epsfbox{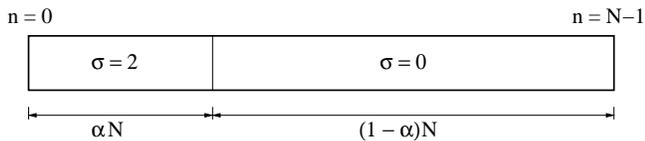}}}\fi
\caption{\small The here considered morphogenetic problem of
boundary formation and scale-independent 
proportion regulation of adjacent domains. 
The target pattern consists of one domain with a fraction of 
$\alpha N$ cells in state $\sigma_i =2$, 
a boundary state $\sigma_i =1$ at the position $\alpha N$, 
and $\sigma_i = 0$ for the remaining domain. }
\end{figure}
A basic means of communication between cells is local information transfer 
between neighbor cells through chemical agents, combined with 
information processing of a cell-internal gene network. Let us include 
local information transfer in the model, similar to the observed mechanism of 
direct contact induction in animals \cite{RansickDavidson1989,Gilbert97,SoleSalazar02}.  
An interesting question is whether global positional information could be an outcome of local information transfer,
in particular, when no concentration-dependent gene activation based on macroscopic gradients is involved. 
As only further requirement, the up-down-symmetry of the body axis has to be broken locally in that case, e.g.,  
by a spatially asymmetric receptor distribution \cite{Marciniac03}.

To define a model system that performs the above task, consider a one-dimensional stochastic 
cellular automaton \cite{Wolfram84} with parallel update. $N$ cells are arranged
on a one-dimensional grid, and each cell is labeled uniquely with an index $i \in \{0,1,...,N-1\}$.
Each cell can take $n$ possible states $\sigma_{i} \in \{0,1,..,n\}$. The state $\sigma_i(t)$ of cell $i$
is a function of its own state $\sigma_i(t-1)$ and of its nearest neighbor's states  $\sigma_{i-1}(t-1)$  
and $\sigma_{i+1}(t-1)$ at time $t-1$, i.e. 
\begin{equation}
\sigma_i(t) = f( \sigma_{i-1}(t-1),  \sigma_i(t-1), \sigma_{i+1}(t-1) ) 
\end{equation}
with $f: \{0,1,...,n\}^3  \mapsto \{0,1,...,n\}$ (cellular automaton with {\em neighborhood 3}).
At the system boundaries, for simplicity, we choose a discrete analogue of zero flux boundary 
conditions, i.e. we set $\sigma_{-1} = \sigma_{N} = const. = 0$. Other choices,
e.g. asymmetric boundaries with cell update depending only on the inner neighbor cell, lead
to similar results.

In the following, let us study the morphogenetic problem formulated above for $\alpha = 0.3$.
We searched for possible solutions in rule space by the aid of genetic algorithms (for details
see \cite{RohlfBornFluct03}). The rule
table of the best solution found is shown in Table I. Fig.\ 2 demonstrates the dynamics.  
\begin{table}
\begin{tabular}{|c|c|c||c|c|c||c|c|c|}\hline
R & $\vec{\sigma}(t-1)$  & $\sigma_i(t)$ & 
R & $\vec{\sigma}(t-1)$ & $\sigma_i(t)$ &
R & $\vec{\sigma}(t-1)$ &  $\sigma_i(t)$ \\ \hline
0 & (0,0,0) & 0 & 9 & (1,0,0) & 0 &  18 & (2,0,0) & 0\\ \hline
1 & (0,0,1) & 2 & 10 & (1,0,1) & 1 &  19 & (2,0,1) & 0\\ \hline
2 & (0,0,2) & 1 & 11 & (1,0,2) & 1 &  20 & (2,0,2) & 0   \\ \hline
3 & (0,1,0) & 0 & 12 & (1,1,0) & 0 &  21 & (2,1,0) & 1 \\ \hline
4 & (0,1,1) & 2 & 13 & (1,1,1) & 1 &  22 & (2,1,1) & 2 \\ \hline
5 & (0,1,2) & 2 & 14 & (1,1,2) & 2 & 23 & (2,1,2) & 2  \\ \hline
6 & (0,2,0) & 1 & 15 & (1,2,0) & 0 & 24 & (2,2,0) & 1  \\ \hline
7 & (0,2,1) & 2 & 16 & (1,2,1) & 0 & 25 & (2,2,1) & 2   \\ \hline
8 & (0,2,2) & 2 & 17 & (1,2,2) & 1 & 26 & (2,2,2) & 2  \\ \hline
\end{tabular}
\caption{\small Rule table of the 3-state cellular automaton. 
First column: rule table index $R$. 
Second column:  input states $\vec{\sigma}=(\sigma_{i-1},\sigma_i,\sigma_{i+1})$ at time $t-1$. 
Third column: corresponding output states at time $t$.}
\end{table}
\begin{figure}[tbp]
\let\picnaturalsize=N
\def\picsize{85mm}
\def\picfilename{ca11final.eps}
\ifx\nopictures Y\else{\ifx\epsfloaded Y\else\input epsf \fi 
\let\epsfloaded=Y
\centerline{\ifx\picnaturalsize N\epsfxsize \picsize\fi
\epsfbox{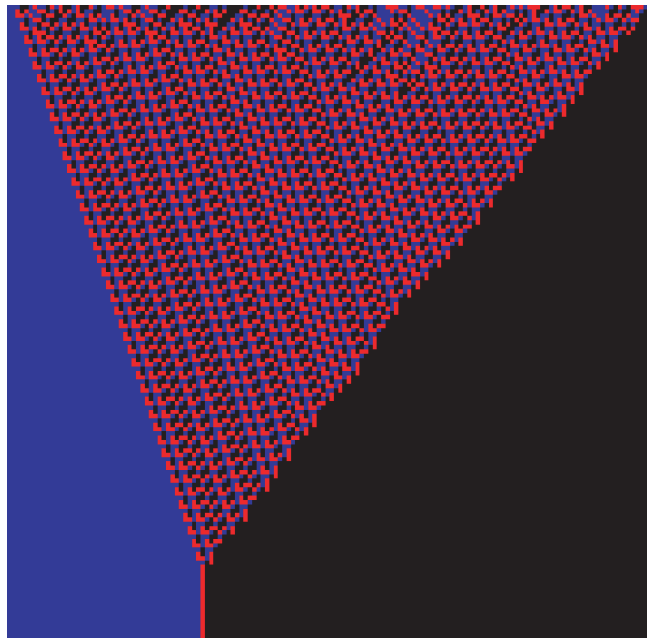}}}\fi
\caption{\small A typical dynamical run for the cellular automata model, time runs from top to bottom. 
Starting from a random initial configuration, cells reorganize into an ordered expression pattern 
corresponding to two asymmetric domains and a sharp domain boundary (deterministic dynamics,
system size $N = 150$, color code of cell states: black $\sigma_i = 0$, red $\sigma_i = 1$, blue $\sigma_i = 2$). }
\end{figure} 
\begin{figure}[tbp] 
\let\picnaturalsize=N
\def\picsize{85mm}
\def\picfilename{ndepnew.eps}
\ifx\nopictures Y\else{\ifx\epsfloaded Y\else\input epsf \fi
\let\epsfloaded=Y
\centerline{\ifx\picnaturalsize N\epsfxsize \picsize\fi
\epsfbox{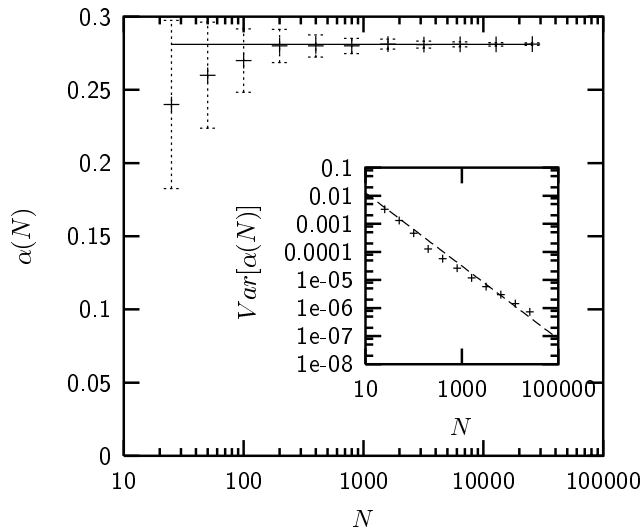}}}\fi
\caption{\small Finite size scaling of relative domain size
$\alpha$ in equilibrium. In the limit of large
system size, $\alpha$ converges to a fixed value $\alpha_{\infty} = 0.281 \pm 0.001$
(corresponding to line fit). Inset: finite size scaling
of the variance $Var[\alpha(N)]$, fluctuations vanish with a 
power $-1.3$ of the system size.}
\end{figure}
Starting from a random initial configuration, the pattern self-organizes towards the target 
pattern within a finite number of updates. The finite size scaling of the self-organized relative 
domain size $\alpha$ as a function of the number of cells $N$ is shown in Fig.\ 3. 
In the limit of large system size, the boundary becomes sharp with $\alpha$ converging towards 
$\alpha_{\infty} = 0.281 \pm 0.001$.
The variance of $\alpha$ vanishes with a power of $N$, i.e. the relative
size of fluctuations induced by different initial conditions becomes
arbitrarily small with increasing system size. The pattern
self-organization in this system is, therefore, quite robust. 
The main mechanism leading
to stabilization at $\alpha_{\infty} = 0.281$ is a modulation of
the traveling velocity of the right phase boundary in Fig.\ 2 due to
particle interactions. On average, the boundary moves
ca.\ one cell to the left per update step, whereas the left boundary
moves one cell to the right every third update step. 
An intuitive concept for the dynamics of cellular automata 
phase boundaries views boundaries as moving particles. 
This so-called ``particle computation'' describes the phenomenology 
of complex cellular automata in terms of these soliton-like excitations \cite{Das94}.  

Let us now study the dynamics of the system under noise. For this purpose,
we introduce stochastic update errors with probability $p$ per cell,
leading to an average error rate $r_e = p\,N$. In numerical simulations,
we now have to apply a statistical method to measure the boundary position
in order to get conclusive results also for high $p$: starting at $i=0$,
we move a measuring frame of $w$ cells to the right and measure
the fraction $z$ of cells with $\sigma_i = 2$ within the frame. The algorithm
stops at some $i$ when $z$ drops below $1/2$ and the boundary position is 
defined as $i + w/2$. It is easy to see that, for not too high $p$, there are only
two different ``particles'' (i.e. state perturbations
moving through the homogeneous phases), as shown in Fig.\ 4. 
In the following,
these particles are called $\Gamma$ and $\Delta$. The  $\Gamma$
particle is started in the $\sigma_2$ phase by a stochastic error $\sigma_i = 2 
 \rightarrow \sigma_i \ne 2$ at some $i < \alpha N$, moves to the right and,
when reaching the domain boundary, readjusts it
two cells to the left of its original position. The $\Delta$ particle is started in the $\sigma_0$
phase by stochastic errors $\sigma_i = 0 \rightarrow \sigma_i \ne 0$ at
some $i > \alpha N$ and moves to the left. Interaction with the interface
boundary readjusts it one cell to the right. 
\begin{figure}[tbp]
\let\picnaturalsize=N
\def\picsize{85mm}
\def\picfilename{noisecompnewfinal.eps}
\ifx\nopictures Y\else{\ifx\epsfloaded Y\else\input epsf \fi
\let\epsfloaded=Y
\centerline{\ifx\picnaturalsize N\epsfxsize \picsize\fi
\epsfbox{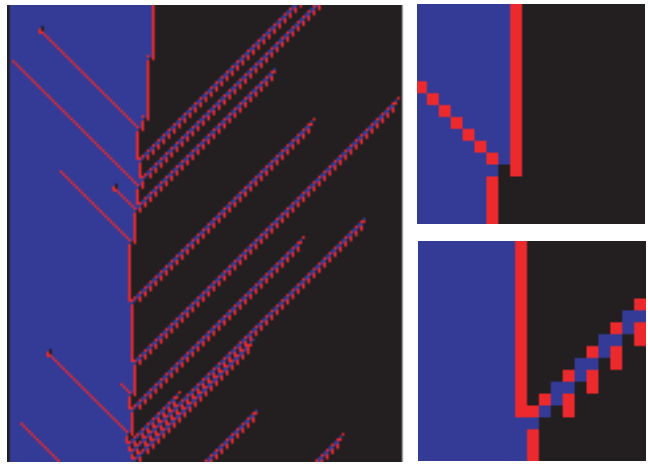}}}\fi
\caption{\small Left panel: Cellular automata dynamics under noise (error rate $r_e = 0.005$). 
Stochastic errors produce two different kinds of quasi-particle excitations, see right 
panels. Top: Interaction of the $\Gamma$-particle with the boundary readjusts the 
boundary two cells to the left. Bottom: $\Delta$-particle moves boundary one cell to the right. }
\end{figure}
Thus we find that the average position $\alpha^*$ of the
boundary is given by the rate equation
\begin{equation} 
2\alpha^* r_e = (1 - \alpha^*) r_e,
\end{equation} 
i.e. $\alpha^* = 1/3$. Interestingly, for not too high error rates $r_e$,  $\alpha^*$
is independent from $r_e$ and thus from $p$. Eqn.\ (2) also implies that the system
undergoes a first order phase transition with respect to  $\alpha^*$ at $p = 0$; numerical evidence
for this conclusion is given in \cite{RohlfBornFluct03}. Fluctuations of $\alpha$ around $\alpha^*$ are Gaussian distributed
with variance vanishing $\sim N^{-1}$ \cite{RohlfBornFluct03}.
The solution  $\alpha^* = 1/3$ is stable only for $0 < r_e \le 1/2$.
\begin{figure}[bp]
\let\picnaturalsize=N
\def\picsize{85mm}
\def\picfilename{phasenew.eps}
\ifx\nopictures Y\else{\ifx\epsfloaded Y\else\input epsf \fi
\let\epsfloaded=Y
\centerline{\ifx\picnaturalsize N\epsfxsize \picsize\fi
\epsfbox{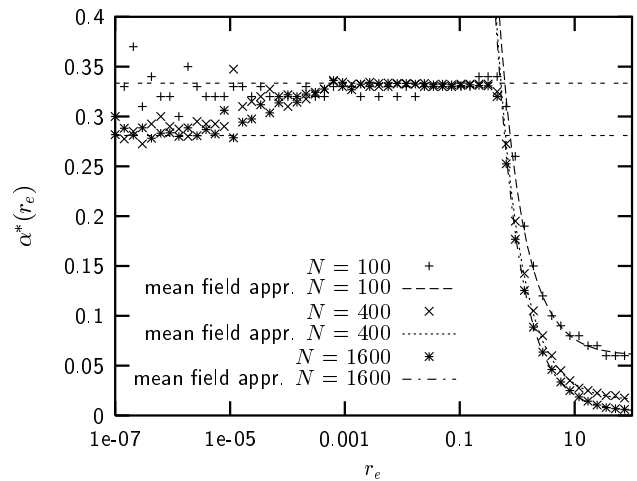}}}\fi
\caption{\small Average boundary position $\alpha^*$ as an order parameter in the presence of noise. 
Transitions are shown as a function of the error rate $r_e$. Numerics averaged over $200$ initial conditions with
$2\cdot10^6$ updates each. Dashed curves: mean field approximation given by Eqn.\ (3). Horizontal 
dashed lines: unperturbed solution $\alpha^* = 0.281$ and solution under noise $\alpha^* = 1/3$. }
\end{figure}
As shown in Fig.\ 4, the interaction of a  $\Gamma$ particle with the boundary
needs only one update time step, whereas the boundary readjustment following a $\Delta$ particle 
interaction takes three update time steps. Therefore, the term on the right hand side of Eqn.\ (2), 
the flow rate of $\Delta$ particles at the boundary will saturate at $1/3$ for large $r_e$, leading to
$2\alpha^* r_e = 1/3$ with the solution 
\begin{equation} \alpha^*  = \frac{1}{6}\,\, r_e^{-1} + \Theta(N) \end{equation} 
for $r_e > 1/2$. Hence, there is a crossover from the solution $\alpha^* = 1/3$ to another solution vanishing
with $r_e^{-1}$ around  $r_e = 1/2$ (Fig.\ 5).
The finite size scaling term $\Theta(N)$ can be estimated from the following
consideration: for $p \rightarrow 1$, the average domain size created by ``pure chance''
is given by $\alpha^* = N^{-1} \,\sum_{n=0}^{N} (1/3)^n \cdot n \approx (3/4)\,N^{-1}$. If the measuring
window has size $w$, we obtain $\Theta(N) \approx (3/4)\,w\,N^{-1}$.

In a biological organism, a pattern has to be robust not only with respect
to dynamical noise, but also with respect to ``mechanical'' perturbations. In \emph{Hydra}, for example, 
there is a steady flow of cells directed towards the animal's head and foot, due to continued proliferation
of stem cells. The stationary pattern of gene activity
is maintained is spite of the cell flow. Let us now study the robustness of the model with respect to this type of perturbation.
Consider a constant cell flow with rate $r_f$ directed towards the left system boundary. 
In Eqn.\ (2), we now get an additional drift term $r_f$ on the left hand side:
$2\alpha^*r_e +  r_f =(1 - \alpha^*)r_e$,
with the solution
\begin{equation}
\alpha^*=  \left\{  
\begin{array}{cccc} \frac{1}{3}\left ( 1 - \frac{r_f}{r_e}\right )  \quad &\mbox{if}&  r_e \ge r_f \quad &\mbox{and}
\quad r_e \le 1/2  \\ 
 0     \quad  &\mbox{if}&  r_e < r_f &
 \end{array}
 \right.
\end{equation}
with $\alpha^*$ exhibiting a second order phase transition at the critical value $r_e^{crit} = r_f$. 
Below $r_e^{crit}$, the domain size $\alpha^*$ vanishes, and above $r_e^{crit}$ it grows until it reaches
the value $\alpha^*_{max}=1/3$ of the system without cell flow. The second order phase transition 
at $r_e^{crit}$ bears some similarity with error catastrophes in models of viral evolution \cite{Kamp}.
\begin{figure}[tbp]
\let\picnaturalsize=N
\def\picsize{85mm}
\def\picfilename{cf.eps}
\ifx\nopictures Y\else{\ifx\epsfloaded Y\else\input epsf \fi
\let\epsfloaded=Y
\centerline{\ifx\picnaturalsize N\epsfxsize \picsize\fi
\epsfbox{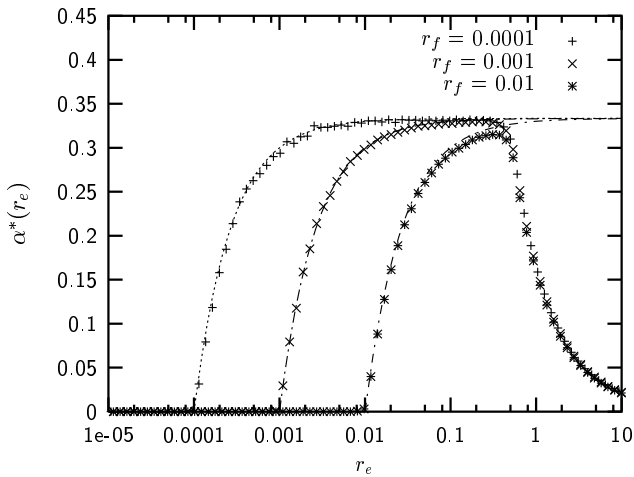}}}\fi
\caption{\small Average boundary position $\alpha^*$ as a function of error rate 
$r_e$ for different cell flow rates $r_f$.
Dashed curves: corresponding solutions of Eqn.\ (4).
Note that a minimum error rate is necessary to regulate the boundary in the presence of cell flow.}
\end{figure}
Fig.\ 6 compares the numerical results 
with the mean field approximation of Eqn.\ (4). In numerical simulations, cell flow was realized by application
of the translation operator $\Theta\,\sigma_i := \sigma_{i+1}$ to all cells with $0 \le i < N-1$ every $r_f^{-1}$ time steps
and leaving $\sigma_{N-1}$ unchanged.
Note that stochastic errors in dynamical updates for $r_e > r_f$ indeed \emph{stabilize} the global pattern
against the mechanical stress of directed cell flow.

To summarize, we considered a problem of pattern formation motivated by animal morphogenesis in a non-traditional setting. 
Accurate regulation of position information, exhibiting proportional scaling with system size, 
and robust {\em de novo} pattern formation from random conditions have been obtained
with a mechanism based on local information transfer rather than the Turing instability. 
Non-local information is transmitted 
through soliton-like quasi-particles instead of long-range gradients, and fine-tuning of parameters is not needed. 
Noise contributes to the stability by generating quasi-particles that control the pattern. We observe considerable 
stability also under cell flow. A first order phase transition is observed for vanishing noise and a second order phase transition 
at increasing cell flow. The pattern formation mechanism studied here is very general and not limited to cellular automata. 
In particular, implementations as regulatory networks work as well \cite{RohlfBornFluct03}, and do not differ in complexity from regulatory circuits 
observed in the cell \cite{BoschDevelGenes}. With this work we hope to inspire new approaches to biological pattern formation 
and to properties of non-equilibrium systems. 

{\bf Acknowledgements} 
We thank T.C.G.\ Bosch, D.\ Drasdo, T.W.\ Holstein, K.\ Klemm, and U.\ Technau  
for inspiring discussions. T.R.\ acknowledges financial support from 
the Studienstiftung des deutschen Volkes (German National Merit Foundation).

\end{document}